\documentclass{aa}  

\usepackage{graphicx}
\usepackage[varg]{txfonts}
\usepackage{xcolor}
\usepackage[colorlinks=true,linkcolor=blue,citecolor=blue,urlcolor=blue]{hyperref}
\usepackage{txfonts}
\usepackage{placeins}
\usepackage{float}
\usepackage[normalem]{ulem}
\usepackage{blindtext}
\usepackage{multirow}
\usepackage[para]{threeparttable}
\bibpunct{(}{)}{;}{a}{}{,}

\begin{document} 

\title{The TNG50-SKIRT Atlas: \\
Multi-wavelength nonparametric galaxy morphology}

\titlerunning{TNG50-SKIRT Atlas}

\author{%
Sena~Bokona~Tulu\inst{\ref{Jimma},\ref{UGent}}\thanks{\email{sena.bokona@ugent.be}}
\and
Maarten~Baes\inst{\ref{UGent}}
\and
Angelos~Nersesian\inst{\ref{UGent},\ref{ULiege}}
\and
Tolu~Biressa\inst{\ref{Jimma}}
\and
Vicente~Rodriguez-Gomez\inst{\ref{UNAM}}
\and
Andrea~Gebek \inst{\ref{UGent}}
\and
Marco~Martorano \inst{\ref{UGent}}
\and
Abdissa~Tassama~Emana \inst{\ref{Jimma},\ref{UGent}}
}

\institute{%
Physics Department, College of Natural Sciences, Jimma University, PO Box 378, Jimma, Ethiopia
\label{Jimma}
\and
Department of Physics and Astronomy, Universiteit Gent, Proeftuinstraat 86 N3, B-9000 Ghent, Belgium
\label{UGent}
\and
STAR Institute, Quartier Agora -- All\'ee du six Ao\^{u}t 19c, B-4000 Li\`ege, Belgium
\label{ULiege}
\and
Instituto de Radioastronom{\'{\i}}a y Astrof{\'{\i}}sica, Universidad Nacional Aut{\'{o}}noma de M{\'{e}}xico, Morelia, Michoac{\'{a}}n 58089, Mexico
\label{UNAM}
}

\date{\today}

%%%%%%%%%%%%%%%%%%%%%%%%%%%%%%%%%%%%%%%%%%%%%%%%%%%%%

\abstract
{Galaxy morphology is a fundamental property to describe galaxy evolution. However, the observed morphology of a particular galaxy may depend on the observed wavelength.}
{Our aim is to investigate the wavelength dependence and the effect of dust attenuation on nonparametric morphology indicators.}
{We use the TNG50-SKIRT Atlas, an atlas of synthetic UV to near-infrared (NIR) broadband images for a complete stellar-mass-selected sample of 1154 galaxies extracted from the TNG50 cosmological simulation at $z = 0$. For each image, we calculate four nonparametric morphology indicators using the {\tt{StatMorph}} code.}
{We find that the known correlations between the stellar mass and the morphological parameters measured in the optical, together with the Gini--$M_{20}$, concentration--Gini, and concentration--$M_{20}$ planes, are fully consistent with observational data. However, nonparametric morphological indicators change significantly with wavelength and that this wavelength dependence is stronger for disc-dominated than for bulge-dominated galaxies. The wavelength dependence of the morphology of our simulated TNG50 galaxies is consistent with measurements of local galaxies from the SINGS survey. We demonstrate that the effect of dust attenuation on nonparametric morphology indicators is modest across the full galaxy population but can be significant for individual galaxies.} 
{}


\maketitle

%%%%%%%%%%%%%%%%%%%%%%%%%%%%%%%%%%%%%%%%%%%%%%%%%%%%%

\section{Introduction} 
\label{sec:introduction}

Galaxy morphology is fundamental not only for classifying galaxies but also for revealing the physical processes that govern their formation and evolution. Morphology allows for a direct representation of the spatial distribution of stars and dust within galaxies, offering key insights into their evolution and assembly history. Morphology is strongly linked with the galaxy’s local environment \citep{Dressler1980, Gomez2003, Blanton2009, Moustakas2023}, merger history \citep{Lotz2008}, stellar mass \citep{Kauffmann2003}, and star-formation history \citep{Kauffmann2003, Baldry2004}. Morphological classification and structural analysis of galaxies, based on their visible shapes, light distribution, and other physical properties, remain powerful diagnostics in observational galaxy studies \citep{Lotz2004}.

There are different techniques to measure the galaxy morphology. The traditional method is based on visual classification \citep{deVaucouleurs1959, Buta2015} and now lives on in the form of large citizen-science projects such as Galaxy Zoo \citep{Dickinson2018} or machine-learning approaches \citep{Sreejith2018, Reza2021}. A more quantitative approach is based on parameters derived from parametric fits to surface brightness profiles, typically in the form of single or multiple-component S\'ersic fits \citep{Gadotti2009, Bottrell2017, Quilley2023, EuclidCollaborationQuilley2026}. Another possibility is the use of nonparametric morphology indicators \citep{Lotz2008, Snyder2015, RodriguezGomez2019, Bignone2020}. Nonparametric morphologies are a robust method for quantifying the morphologies of galaxies. They do not assume an underlying model for the light distribution of a galaxy and are therefore well suited to any type of galaxy regardless of wavelength or redshift \citep{Conselice2003, Lotz2004, Lotz2008, Lotz2010, Freeman2013}. Furthermore, nonparametric morphology can be applied to images outside the traditional optical regime: recent applications of nonparametric morphology over the entire UV to radio regime include \citet{MunozMateos2009}, \citet{Holwerda2011}, \citet{Baes2020c}, \citet{Kapoor2021}, \citet{Camps2022}, and \citet{Gebek2023}. 
 
Numerical simulations can play a crucial role in exploring the origin of galaxy morphologies. Unlike observations, in simulations, the full three-dimensional distribution of stars, gas, and dust is known, and the effects of different sources and processes can be separated to quantify their impact on morphological indicators. Modern cosmological hydrodynamical simulations, such as EAGLE \citep{Schaye2015}, SIMBA \citep{Dave2019}, TNG100 \citep{Pillepich2018b, Marinacci2018, Nelson2018, Springel2018, Naiman2018}, TNG50 \citep{Nelson2019b, Pillepich2019} or NewHorizon \citep{Dubois2021}, have led to significant advances in our understanding of galaxy formation. A key test for these cosmological simulations is whether they can produce galaxies with realistic morphologies consistent with those of observed galaxies. To this end, it is essential to establish a robust mapping between simulation results and morphological diagnostics that can be directly compared to those derived from observations. One powerful approach is the generation of synthetic galaxy images from simulations, enabling morphology to be analysed using the same tools applied to observational data \citep[e.g.,][]{Lotz2008, Jonsson2006, Bignone2017, Trayford2017, RodriguezGomez2019}. Several codes, with different focus and different levels of sophistication, have been developed to generate synthetic images for simulated galaxies, including {\tt{SUNRISE}} \citep{Jonsson2006}, {\tt{SKIRT}} \citep{Camps2015a, Camps2020}, {\tt{POWDERDAY}} \citep{Narayanan2021}, {\tt{SYNTHESIZER}} \citep{Lovell2025, Roper2026}, or {\tt{GalaxyGenius}} \citep{Zhou2025}. 

Nonparametric morphology is particularly valuable in comparing simulated and observed galaxies, as it enables direct testing of how well simulations reproduce structural diversity. Several standard nonparametric indicators are now widely used, including the CAS (concentration, asymmetry, and smoothness) indices \citep{Conselice2003}, and the Gini and $M_{20}$ coefficients \citep{Abraham2003, Lotz2004}. Several observational studies have applied nonparametric morphology metrics across a range of wavelengths. Nonparametric morphology has been used extensively to quantify the structure of the galaxy in the ultraviolet (UV), optical, and near-infrared (NIR) \citep[e.g.,][]{Lotz2006, HuertasCompany2009, Conselice2009, Conselice2014, Holwerda2011, RodriguezGomez2019}. \citet{Nersesian2023} applied these metrics to optical IFU data with the goal of measuring the galaxy morphology for different emission lines. \citet{MunozMateos2009} and \citet{Baes2020c} extended the analysis to the far-infrared (FIR) and submm regimes using \textit{Spitzer} and {\textit{Herschel}} images, respectively. Nonparametric morphology has also been applied to CO and H{\sc{i}} maps \citep{Holwerda2011, Davis2022, Gebek2023}. 

Many recent studies have applied these metrics to synthetic images of simulated galaxies to evaluate the effectiveness of morphology as a diagnostic of galaxy evolution \citep{Snyder2015, RodriguezGomez2019, Bignone2020, GuzmanOrtega2023}. For instance, \citet{Bignone2020} studied the optical $g$-band morphology of approximately 3600 galaxies in the EAGLE simulation, while \citet{RodriguezGomez2019} investigated the optical $i$-band morphology of more than 12,000 and 14,000 galaxies in TNG100 and original Illustris, respectively. Recently, \citet{GuzmanOrtega2023} generated over 16,000 synthetic images of TNG50 galaxies in the $r$-band to investigate the connection between galaxy mergers and optical morphology. Overall, these studies found that the optical morphologies of simulated galaxies are in good agreement with observations. Although these studies looked at large samples, they are limited to the optical wavelength range. On the other hand, \citet{Kapoor2021} and \citet{Camps2022} analysed a limited sample of galaxies from Auriga and ARTEMIS simulations to show the change in morphological indicators as a function of wavelength from ultraviolet (UV) to submm wavelengths. They found that the morphology of the Auriga and ARTEMIS galaxy samples varies with wavelength and follows the same overall trends as the comparable DustPedia galaxy sample, with some notable discrepancies. \citet{Gebek2023} also presented nonparametric morphologies of the H{\sc{i}} emission at 21~cm of 1130 TNG50 galaxies and found that the TNG50 HI maps are less concentrated than those of observed galaxies from the WHISP survey. Although extremely interesting tests, all of these studies are limited in wavelength coverage or sample size.

In this work, we address these limitations by analysing a large sample of galaxies from the TNG50 simulation using synthetic images from the TNG50-SKIRT Atlas \citep{Baes2024a}. The TNG50-SKIRT Atlas is a large multi-wavelength synthetic image atlas based on simulated galaxies extracted from the TNG50 cosmological hydrodynamical simulation (high spatial resolution combined with large volume) using {\tt{SKIRT}}. This dataset provides both dust-free and dust-attenuated images across a broad wavelength range, from UV to NIR. We used this unique dataset to investigate how nonparametric morphology varies with wavelength and dust content and to compare TNG50 morphologies with other simulations and observations. 

The paper is organized as follows: Section~\ref{sec:data} describes our sample selection, and Section~\ref{sec:methods} describes the methods. In Section~\ref{sec:optical_morphology}, we outline the optical morphology of these TNG50 galaxies and some previous work. The main findings are described in Sections~\ref{sec:npd_vs_lambda} and \ref{sec:dust_effect_on_npd}. Subsequently, Section~\ref{sec:summary_and_conclusions} presents the conclusion of our work.

%%%%%%%%%%%%%%%%%%%%%%%%%%%%%%%%%%%%%%%%%%%%%%%%%%%%%

\section{Data and samples}\label{sec:data}

\subsection{The TNG50 cosmological hydrodynamical simulation}
\label{subsec:tng50_simulations}

The TNG50 simulation \citep{Nelson2019b, Pillepich2019} is the most computationally demanding and highest resolution realization of the cosmological magnetohydrodynamical simulation suite of the IllustrisTNG project. It evolves dark matter (DM), gas, stars, black holes, and magnetic fields within a uniform periodic-boundary cube of 51.7 comoving Mpc on a side, containing $2 \times 2160^3$ total initial resolution elements: half DM particles and half Voronoi gas cells, in addition to an equal number of Monte Carlo tracer particles. A defining feature of TNG50 in comparison to other large-volume cosmological hydrodynamical simulations is its high numerical resolution. In TNG50, the DM particle mass is $4.5\times10^{5}~\mathrm{M}_{\odot}$ , while the mean baryonic gas mass resolution is $8.5\times10^{4}~\mathrm{M}_{\odot}$ and the average cell size in the star-forming regions of galaxies is 70--140~pc. This combination of large volume and high mass and spatial resolution is ideal to study the structure and substructures of galaxies on a population level. 

\subsection{The TNG50-SKIRT Atlas}
\label{subsec:tsa}

{\tt{SKIRT}} \citep{Camps2015a, Camps2020} is a three-dimensional Monte Carlo radiative transfer code, developed and maintained by the astronomy group at UGent. Originally developed to model the effects of dust in galaxies \citep{Baes2003, Baes2011b}, it has since evolved into a versatile tool for general radiative transfer applications. Physical ingredients within {\tt{SKIRT}} include absorption, multiple scattering, and stochastic emission \citep{Camps2015a}; polarization caused by scattering or by emission from aligned non-spherical dust grains \citep{Peest2017, Vandenbroucke2021}; Ly$\alpha$ resonant line scattering \citep{Camps2021}; absorption and emission at rotational or electronic transitions for selected ions, atoms, and molecules \citep{Gebek2023, Matsumoto2023}; photoabsorption, fluorescence, and scattering at X-ray wavelengths \citep{VanderMeulen2023, VanderMeulen2024}. The code is equipped with a large library of stellar population libraries, physical dust models, geometrical models \citet{Baes2015}. Various 3D grid structures with efficient photon traversal methods are available to discretize the medium, making the code efficient even for complex 3D geometries \citep{Camps2013, Saftly2013, Saftly2014, Lauwers2024}.
       
The TNG50-SKIRT Atlas \citep[TSA,][]{Baes2024a} is a synthetic image database for galaxies extracted from the $z = 0$ snapshot of the TNG50 simulation. It includes all galaxies with a total stellar mass in the range $10^{9.8}$ to $10^{12}~{\text{M}}_\odot$, resulting in a sample of 1154 galaxies. The lower limit of this range was chosen to ensure that the simulated galaxies are still sufficiently well resolved; below this limit, galaxies are increasingly affected by resolution effects, which impact the spatial distribution of stars and dust, and therefore the robustness of morphology measurements. For the current paper, we restrict the sample to galaxies with $M_\star \leqslant10^{11.3}~{\text{M}}_\odot$, in agreement with the comparison samples we use (Sect.~{\ref{sec:optical_morphology}}). The stellar mass range $10^{9.8}$ -- $10^{11.3}~{\text{M}}_\odot$ spans a broad diversity of galaxy types, from low-mass disc-dominated systems to massive bulge-dominated galaxies.

The TSA contains dust-attenuated and dust-free synthetic images in 22 UV--NIR broadband filters\footnote{The first data release, presented in \citet{Baes2024a}, contains images in 18 UV--NIR broadband filters. Additional dust-attenuated and dust-free images in the four {\it{Euclid}} broadband filters are presented in \citet{EuclidCollaborationKovacic2025}.} and four physical parameter maps, for five random observing positions per galaxy. In total, this adds up to 240 images per galaxy. In this work, we limit the analysis to a subset of 12 broadband filters spanning the UV to NIR wavelength range, providing uniform coverage across this domain. More specifically, we use images in the GALEX NUV and FUV bands, the LSST {\it{ugriz}} bands, the 2MASS {\it{JHK$_{\text{s}}$}} bands, and the WISE W1 and W2 bands. Moreover, we only use the images corresponding to the random orientations O1--O4, as the fifth observer position (O5) was designed to be antipodal to O4 \citep{Baes2024a}. All images and parameter maps are defined on a $1600 \times 1600$ pixel grid, with a physical pixel scale of 100~pc, corresponding to a field of view of 160~kpc.

%%%%%%%%%%%%%%%%%%%%%%%%%%%%%%%%%%%%%%%%%%%%%%%%%%%%%

\section{Measuring galaxy morphology with {\tt StatMorph}}
\label{sec:methods}

To compute the nonparametric morphologies of our synthetic galaxy images, we use {\tt StatMorph} \citep{RodriguezGomez2019}, a {\tt python} package for both nonparametric and parametric morphological analysis. It has been widely used in the astronomical community \citep[e.g.,][]{Baes2020c, Nersesian2023, Yao2023, Leste2024, Drigga2025, Tian2025}.

To enable a realistic analysis of nonparametric morphology indicators with {\tt StatMorph}, we preprocessed the TSA images to emulate observational effects. We convolve the images with a Gaussian PSF with a standard deviation of three pixels and we add Gaussian background noise consistent with a peak signal-to-noise ratio (S/N) of 100 at the brightest pixel in the image. Furthermore, each TSA image was cropped to eight times the stellar half-mass radius to focus on the main galaxy structures. Eventually, we generated segmentation maps using the {\tt python} package {\tt photutils} and performed morphology measurements using {\tt StatMorph}. We applied a quality selection in all band analysis. Specifically, we removed all sources with a flag~$> 1$, which is indicative of a poor quality measurement \citep{RodriguezGomez2019}. We also removed sources with a mean S/N ratio per pixel less than two and with a half-mass radius smaller than 2~kpc. The reason is that a low S/N per pixel or insufficient resolution affects the precision of morphological parameter measurements \citep{Lotz2006}. As a result, approximately 75\% of our original sample meets all criteria in both the dust-free and dust-attenuated cases. In total, we consider a set of 4\,329 individual image sets.

We compute nonparametric morphology indicators using the {\tt StatMorph} code. Although this code provides a wide range of diagnostics, in this work we focus on the concentration, Gini, $M_{20}$, and bulge statistic.

The concentration parameter ($C$) is defined as follows \citep{Conselice2003},
\begin{equation}
C = 5 \log_{10}\left(\frac{R_{80}}{R_{20}}\right),
\end{equation} 
where $R_{80}$ is the radius of the circular aperture enclosing 80\% of the total flux and $R_{20}$ is the radius of the aperture enclosing 20\% of the total flux. The total flux is measured within a 1.5 Petrosian radius aperture and the centre of the aperture corresponds to the point that minimizes the asymmetry, as determined by {\tt StatMorph}.

The Gini coefficient ($G$) is a statistical indicator that measures the distribution of light among the pixels that comprise the galaxy image. It is defined as \citep{Abraham2003, Lotz2004}
\begin{equation}
G = \frac{1}{\bar{f}\,n\,(n-1)}
\sum_{i=1}^n (2i-n-1) f_i
\end{equation}
where $n$ is the number of pixels in the galaxy region, $f_i$ is the intensity of pixel $i$ (sorted in increasing order), and $\bar{f}$ is the mean pixel intensity. If $G=0$, the light is evenly distributed over all galaxy pixels, while $G=1$ corresponds to the theoretical limit where all flux is concentrated in just one pixel.

The $M_{20}$ parameter is the second-order moment of the brightest regions of the galaxy \citep{Lotz2004}, defined as the region consisting of the pixels, ordered in decreasing flux, that jointly contain 20\% of the total flux. Similarly to the concentration index it is diagnostic for the light concentration, but it is not restricted to the central region: it is sensitive to any bright nucleus, bar, spiral arm, or off-centre clump. The value of $M_{20}$ is always negative.

Finally, the bulge statistic, $F$, is a linear combination of the Gini and $M_{20}$ indices and is defined as five times the distance to the line separating the disc-dominated and bulge-dominated populations in the Gini--$M_{20}$ plane \citep{Snyder2015, RodriguezGomez2019, Bignone2020}. It is a quantitative measure used to classify galaxies based on their bulge prominence, with positive (negative) values corresponding to bulge-dominated (disc-dominated) galaxies \citep{Snyder2015}. 

Although we initially also considered asymmetry and smoothness as morphological indices, we find that these two quantities are not sufficiently robust for the present dataset and are therefore not included in the analysis presented in this work. In practical implementations, both asymmetry and smoothness involve a subtraction of a background term, which can lead to formally negative values when the background contribution exceeds the intrinsic signal of the galaxy \citep[e.g.][]{Conselice2003, Lotz2004, RodriguezGomez2019, Thorp2021}. While such values are in principle allowed, we find that a large fraction of galaxies in our mock images exhibit strongly negative asymmetry and smoothness values.

We have investigated the origin of this behaviour and find that it  is not primarily driven by the adopted background noise prescription. In particular, we repeated the analysis using an alternative background noise model in which the noise level is set by a fixed average signal-to-noise ratio across the image, rather than by the peak signal. This modification does not significantly reduce the occurrence of negative values. We therefore conclude that the measurements of asymmetry and smoothness are strongly affected by the combination of Monte Carlo noise in the {\tt{SKIRT}} radiative transfer images and the high spatial resolution of the data (pixel scale of 100 pc), which enhances pixel-to-pixel fluctuations.

In contrast, the other nonparametric indicators considered in this work ($C$, $G$, $M_{20}$, and $F$) are significantly more stable against noise and yield consistent trends across the sample. We therefore restrict our analysis to these quantities in the remainder of this paper.

%%%%%%%%%%%%%%%%%%%%%%%%%%%%%%%%%%%%%%%%%%%%%%%%%%%%%

\section{Optical morphology}
\label{sec:optical_morphology}

\begin{figure*}[t]
\centering
\includegraphics[width=\textwidth]{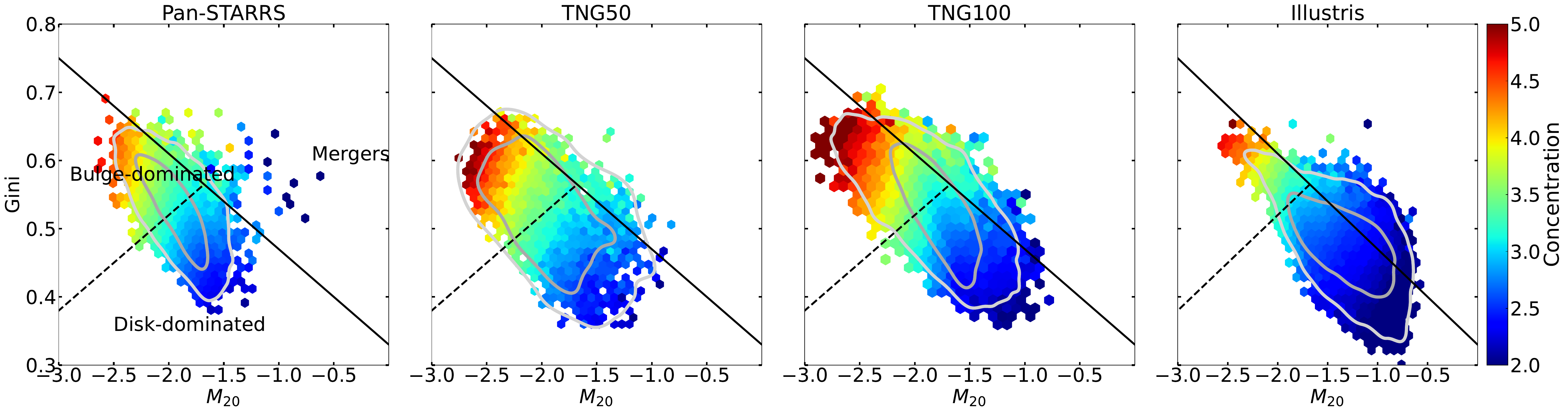}
\caption{
{\it{i}}-band Gini--$M_{20}$ diagram for the different samples considered in this work. From left to right: Pan-STARRS, our TNG50 galaxy sample, TNG100, and original Illustris simulations. The solid and dashed black lines are taken from \citet{Lotz2008}. The solid black line roughly separates isolated galaxies from mergers, whereas the dashed black line divides galaxies into early- and late-types. The dark (inner) and light (outer) grey contours contain 68\% and 95\% of the galaxy distribution, respectively.}
\label{fig:figure_1}
\end{figure*}

The main goals of this paper are to investigate the wavelength dependence and the effects of dust attenuation on nonparametric galaxy morphology indicators. Before we turn to these topics in the next two Sections, we first investigate the morphology of our TNG50 galaxy sample in the optical $i$-band. We compare our results with similar results obtained for the original Illustris simulation \citep{Vogelsberger2014} and the TNG100 simulation \citep{Pillepich2018b, Marinacci2018, Nelson2018, Springel2018, Naiman2018}, and with galaxies observed in the frame of the Pan-STARRS survey \citep{Chambers2018}. The nonparametric morphological parameters for these comparison samples were derived from \citet{RodriguezGomez2019}, and we refer to that study for a full description of the sample and the methodology. 

%%%%%%%%%%%%%%%%%%%%%%%%%%%%%%%%%%%%%%%%%%%%%%%%%%%%%

\subsection{The Gini--$M_{20}$ plane}
\label{subsec:comparison_with_gomez}

The distribution of galaxies in the Gini--$M_{20}$ plane \citep{Lotz2008} is a commonly used way to classify galaxies \citep[e.g.,][]{Snyder2015, RodriguezGomez2019, Bignone2020, Ren2024}. Figure~{\ref{fig:figure_1}} shows the distribution of galaxies in the $i$-band Gini--$M_{20}$ plane for our TNG50 sample as well as results for three comparison samples taken from \citet{RodriguezGomez2019}, who derived nonparametric morphology measurements for these samples using a consistent methodology.
\begin{itemize}
\item 
The Pan-STARRS sample contains observed galaxies from the PS1 Optical Galaxy Survey \citep[POGS,][]{Vinsen2013}, cross-matched with the NASA Sloan Atlas. The galaxies are selected in the redshift range $0.045 < z < 0.054$ and with NASA Sloan Atlas stellar masses in the range $M_\star \approx 10^{9.8}-10^{11.3}~{\text{M}}_\odot$. The total number of galaxies in this sample is 10\,829.
\item The TNG100 sample contains simulated galaxies from the TNG100 simulation, the flagship run of the IllustrisTNG suite. Compared to TNG50, TNG100 covers an eight times larger cosmological volume, at the cost of a poorer resolution (target baryonic mass resolution of $1.4\times10^6~{\text{M}}_\odot$ compared to $8.5\times10^4~{\text{M}}_\odot$ for TNG50). The sample shown in Figure~{\ref{fig:figure_1}} contains the 12\,069 simulated galaxies from the $z=0.0485$ snapshot (snapshot number 95) with total stellar mass in the range $M_\star = 10^{9.8} - 10^{11.3}~{\text{M}}_\odot$. 
\item Finally, the Illustris sample contains simulated galaxies from the original Illustris simulation \citep{Vogelsberger2014}, the predecessor of the IllustrisTNG suite. This simulation has a comparable resolution and volume as TNG100. The sample shown in Figure~{\ref{fig:figure_1}} contains the 14\,314 simulated galaxies with $M_\star = 10^{9.8} - 10^{11.3}~{\text{M}}_\odot$ from the $z=0.0485$ snapshot (snapshot number 131).
\end{itemize}
We note that detailed morphological classifications (e.g. bulge- versus disc-dominated fractions) for these samples are discussed in \citet{RodriguezGomez2019} and are not repeated here.

Each panel in Figure~{\ref{fig:figure_1}} shows the distribution of galaxies in the $i$-band Gini--$M_{20}$ plane, with the colour in each hexbin indicating the median concentration of all galaxies falling on that pixel. The inner and outer contours indicate the 68\% and 95\% percentile regions around the most dense part of the distributions in each data sample. The solid and dashed lines roughly separate the merger candidates (above the solid line) from bulge-dominated (above the dashed line) and disc-dominated (below the dashed line) according to \citet{Lotz2008}.

In general, the distribution of galaxies in the Gini--$M_{20}$ diagram from TNG50 shows a very satisfactory agreement with the PAN-STARRS observations and the TNG100 simulation. In particular, TNG50 galaxies within the inner contour (enclosing 68\% of the sample) exhibit concentration index values ranging from $C \approx 2.5$ to $C \approx 4$, consistent with observational data and results from TNG100. The distribution of Illustris galaxies in the Gini--$M_{20}$ plane is quite different, with many more galaxies having lower Gini indices and higher $M_{20}$ indices than for the other samples. This result was already described in detail by \citet{RodriguezGomez2019}. 

There is one notable difference between the distribution of galaxies in the TNG50, TNG100 and Pan-STARRS samples that deserves some attention. Pan-STARRS galaxies are slightly skewed toward higher Gini coefficients and less negative $M_{20}$ values, placing a larger fraction of them further into the merger-dominated region of the morphological space. However, a non-negligible fraction of galaxies in all samples is located in the merger-dominated region of the Gini--$M_{20}$ plane, and this fraction appears relatively large for a low-redshift galaxy population. This behaviour is likely not driven by a specific dataset, but rather reflects limitations of the empirical division between mergers and non-mergers proposed by \citet{Lotz2008}, which is known to be subject to contamination. In particular, edge-on disc galaxies can occupy the merger region due to projection effects and the presence of dust lanes, which reduce the apparent central concentration and increase the relative importance of off-centre structures. Inspection of the simulated galaxies confirms that a significant fraction of objects classified as mergers based on this criterion are in fact edge-on systems rather than interacting galaxies. This effect is present across all samples and likely explains the relatively large number of galaxies found in the merger-dominated region. Similar behaviour has been noted in previous studies \citep[e.g.,][]{Bignone2017, Nersesian2023}, in agreement with our findings.

%%%%%%%%%%%%%%%%%%%%%%%%%%%%%%%%%%%%%%%%%%%%%%%%%%%%%

\subsection{Relations among various nonparametric morphology indicators}
\label{subsec:npd_relations}

\begin{figure*}[t]
\centering
\includegraphics[width=0.7\textwidth]{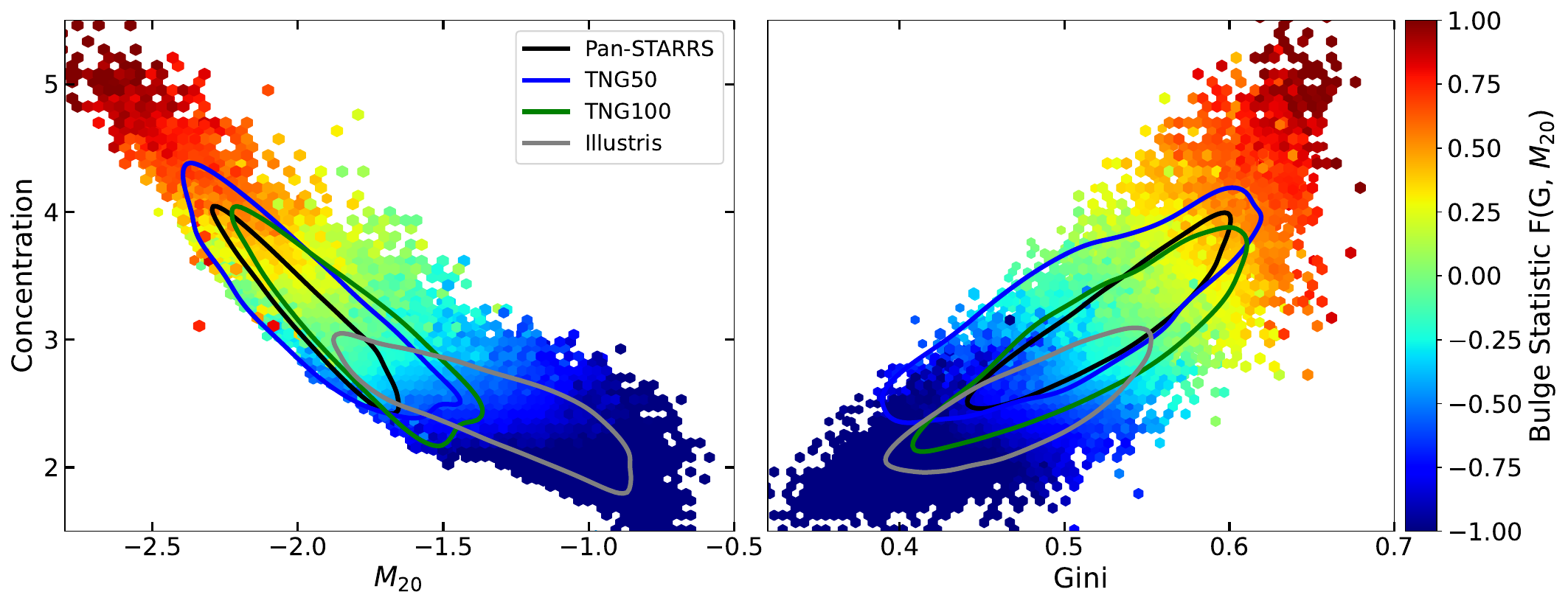}
\caption{Comparison of simulated and observational galaxies in different $i$-band nonparametric morphology indicators: the concentration index as a function of $M_{20}$ and Gini coefficient. The relation between the concentration index and each indicator is coloured according to the median value of the bulge statistic $F$ of the TNG50 galaxies. Higher concentration generally correlates with more bulge-dominated systems (higher $F$). Galaxies with high Gini or low $M_{20}$ values cluster toward higher concentration and bulge statistic, consistent with bulge-dominated morphologies. The bulge statistic increases with concentration and aligns with known morphological trends: high Gini, low-$M_{20}$ galaxies are typically bulge-dominated. The solid contours represent 68\% of the galaxy distribution in their respective samples, as shown in each panel.}
\label{fig:figure_2}
\end{figure*}

The Gini--$M_{20}$ plane already demonstrated a clear correlation between Gini, $M_{20}$, and concentration. Figure~\ref{fig:figure_2} shows two more panels that demonstrate the interdependency of different nonparametric morphology indices. The panels show the distribution of the TNG50 galaxies in the $i$-band concentration--$M_{20}$ and concentration--Gini planes. The hexbins are colour-coded according to the median bulge statistic of TNG50 galaxies, and the contours indicate the 68\% percentile regions of the galaxy distribution in our simulated galaxy sample and the comparison samples. 

The left panel of Fig.~{\ref{fig:figure_2}} shows an anti-correlation of central light concentration and $M_{20}$, in which a more negative $M_{20}$ means more centrally concentrated light. This relation indicates that $M_{20}$ becomes more negative as light is concentrated in the central area. \citet{Lotz2004, Lotz2008} describe a strong anti-correlation relationship between $C$ and $M_{20}$ for non-merging galaxies. Most bulge-dominated galaxies occupy high-concentration and more negative $M_{20}$ regions, while disc-dominated galaxies occupy lower concentration and higher (less negative) $M_{20}$ regions in simulated and observational samples. The distribution of TNG50 galaxies in the concentration-$M_{20}$ plane agrees well with TNG100 and Pan-STARRS samples. Specifically, the 68\% percentile contour of our TNG50 sample shows a similar behaviour to that of TNG100 and observational samples. The Illustris distribution is again relatively far off the other three samples.

The right panel of Fig.~\ref{fig:figure_2} shows a strong positive correlation between concentration and the Gini coefficient. Galaxies with higher concentration generally have higher Gini coefficients \citep{Lotz2004}. Bulge-dominated galaxies exhibit higher concentration and also show higher values of the Gini coefficient and the bulge statistic, whereas disc-dominated galaxies show lower concentration and the Gini coefficient as well as a negative bulge statistic. The TNG50 galaxies agree well with the TNG100 and Pan-STARRS samples, whereas the Illustris galaxies occupy a region in the Gini--concentration plane that corresponds to less concentrated galaxies with lower Gini indices.
  
%%%%%%%%%%%%%%%%%%%%%%%%%%%%%%%%%%%%%%%%%%%%%%%%%%%%%

\subsection{Variation with stellar mass}
\label{subsec:npd_vs_mstar}

\begin{figure*}[t]
\centering
\includegraphics[width=0.8\textwidth]{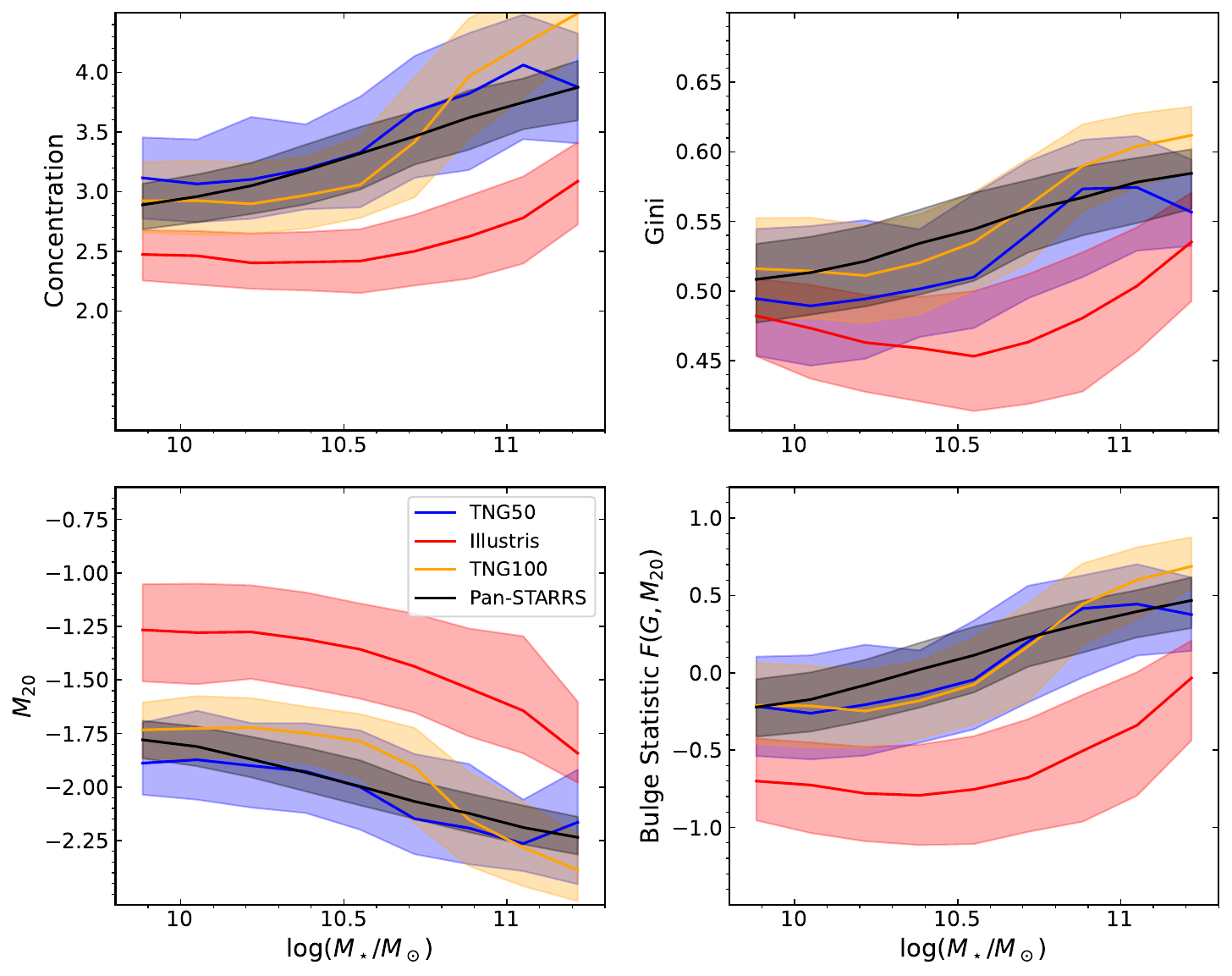}
\hspace*{2em}
\caption{Median trends of nonparametric morphology indicators as a function of stellar mass. The black line shows the median trend for the observational Pan-STARRS sample. The blue line represents the TNG50 simulation, while the yellow and red lines correspond to the TNG100 and original Illustris simulations, respectively, in each panel. The shaded areas around each median line in each panel represent the interquartile range ($25^\mathrm{th}$ to $75^\mathrm{th}$) of the galaxies in that stellar mass bin. The nonparametric morphology indicators from the TNG50 simulation align well with both Pan-STARRS observation and TNG100 simulation.}
\label{fig:figure_3}
\end{figure*}

Figure~\ref{fig:figure_3} compares the four nonparametric morphology indicators as a function of stellar mass for our TNG50 sample and for the comparison samples (Pan-STARRS, TNG100 and Illustris). The black solid lines and gray shaded regions indicate the median and the $25\%$ to $75\%$ percentile range (at a fixed stellar mass) for the Pan-STARRS observations, while the blue, yellow, and red lines and the corresponding shaded regions show the median and interquartile ($25\%$ to $75\%$) range correspond to the TNG50, TNG100, and Illustris cosmological simulations, respectively. 

It is clear that, on average, the nonparametric morphology indicators in the TNG50 galaxies are in very good agreement with the Pan-STARRS data and the TNG100 results over the entire stellar mass range. Concentration, Gini coefficient, and bulge statistics show monotonic increases with stellar mass, whereas $M_{20}$ decreases with stellar mass, in agreement with \citet{RodriguezGomez2019} and \citet{Gong2025}. This indicates that the most massive galaxies are, on average, more bulge-dominated and centrally concentrated. Lower mass galaxies are more irregular, clumpy, and disc-dominated. 

%%%%%%%%%%%%%%%%%%%%%%%%%%%%%%%%%%%%%%%%%%%%%%%%%%%%%

\section{Wavelength dependence of nonparametric morphology indicators}
\label{sec:npd_vs_lambda}

The dependence of nonparametric morphological indicators on wavelength plays a key role in galaxy classification and in understanding the spatial distribution of stellar populations \citep{Bendo2007, Nersesian2023}. Most previous studies in this area have focused on the optical range \citep{RodriguezGomez2019, Bignone2020}, although some authors have recently considered applications in different wavelength regimes \citep{MunozMateos2009, Holwerda2011, Kelvin2012, Baes2020c, Kapoor2021, Camps2022, Nersesian2023}. In this section, we systematically compare the wavelength dependence of the four morphology indicators across the UV--NIR range.

%%%%%%%%%%%%%%%%%%%%%%%%%%%%%%%%%%%%%%%%%%%%%%%%%%%%%

\subsection{Wavelength dependence of individual indicators}
\label{subsec:npd_vs_lambda}

\begin{figure*}[t]
\centering
\includegraphics[width=0.8\textwidth]{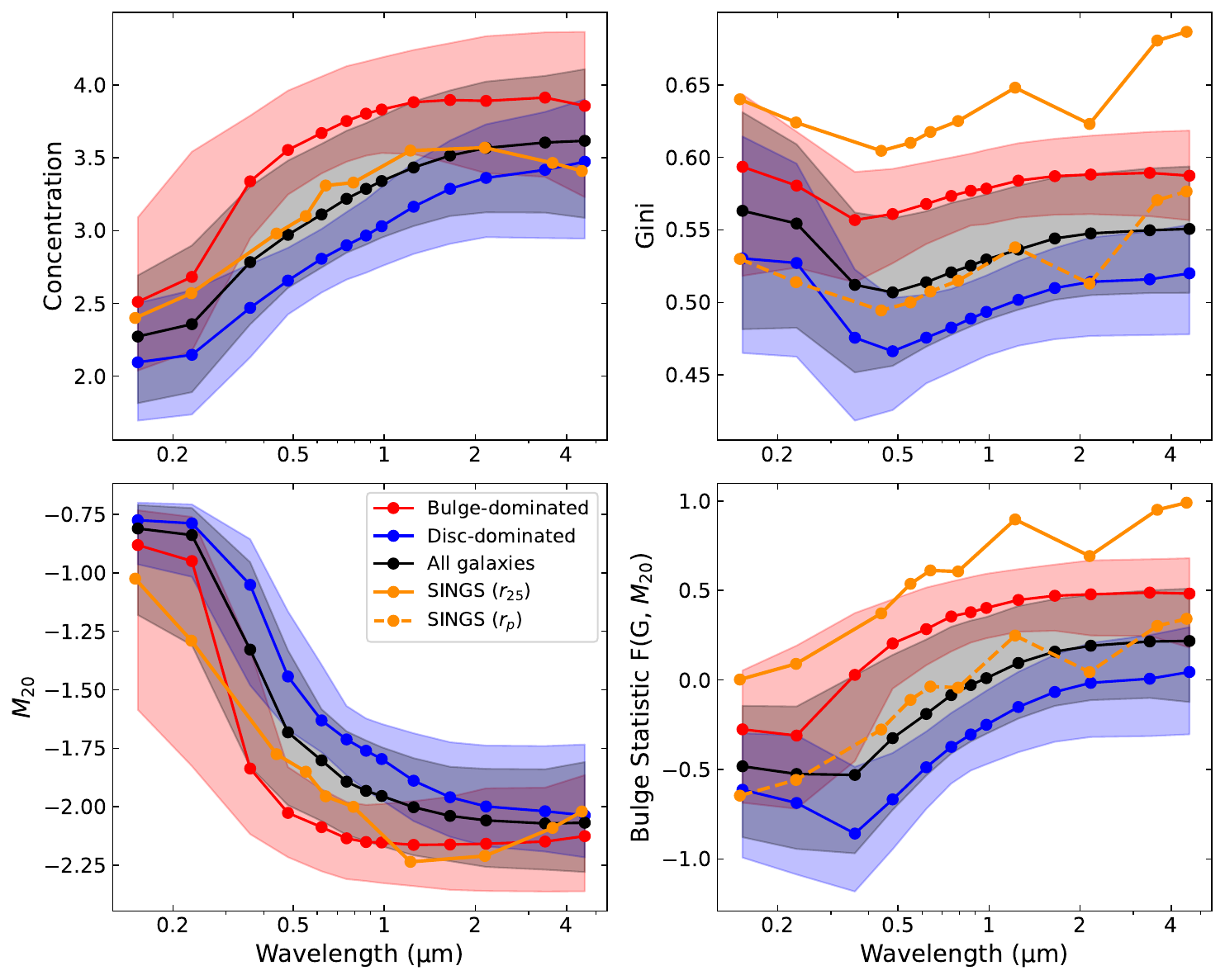}
\hspace*{2em}
\caption{Variation of nonparametric morphology indicators as function of wavelength for bulge- and disc-dominated galaxies. Galaxies are classified based on their bulge statistic $F$, with red lines representing bulge-dominated galaxies ($F > 0$) and blue lines for disc-dominated ones ($F < 0$), whereas the black line indicates the median of all galaxy populations. The $x$-axis represents wavelengths spanning from the FUV to the NIR range, and each panel shows the median value and interquartile range (25--75\%) of a nonparametric morphology indicator as a function of wavelength. For the Gini and bulge statistic, we show two different curves for the SINGS galaxies: the solid line corresponds to measurements within the optical radius, the dashed line to measurements within the Petrosian radius. The concentration index is not tied to a fixed aperture such as $R_{25}$, as it is defined in terms of radii enclosing fixed fractions of the total flux, while for $M_{20}$ the relative difference between aperture definitions is only a few percent \citep{MunozMateos2009}.}
\label{fig:figure_4}
\end{figure*}

The black lines in Figure~\ref{fig:figure_4} show the median trends of each nonparametric morphological indicator as a function of wavelength, for the entire population of the TNG50 galaxy. The red and blue lines show the same results, but for the subpopulation of bulge-dominated and disc-dominated galaxies, respectively. The subdivision of the entire sample into these two populations is based on the $i$-band bulge statistic: galaxies with $F \geqslant 0$ are bulge-dominated galaxies, whereas galaxies with $F<0$ are disc-dominated. 

The solid and dashed orange lines in the different panels correspond to the median trends for the galaxies in the SINGS survey \citep{Kennicutt2003}, measured within different apertures. The solid orange lines indicate the nonparametric indicators measured within the elliptical aperture $R_{25}$, while the dashed orange lines correspond to those computed within the Petrosian radius $r_{\text{p}}$. The SINGS observational sample covers a broad wavelength range from the ultraviolet to the far-infrared, making it well-suited for investigating the wavelength dependence of nonparametric morphology indicators, unlike the Pan-STARRS observational sample, which is limited to a single broadband filter. For the 75 nearby galaxies from this sample, \citet{MunozMateos2009} calculated the concentration, Gini, and $M_{20}$ from UV to far-infrared wavelengths.

Focusing on the top-left panel of Fig.~\ref{fig:figure_4}, we note that the concentration index increases with wavelength from UV--NIR in TNG50 galaxies. Bulge-dominated galaxies consistently show concentration values greater than 2.5 across all wavelengths, with only minor variation in the optical--NIR regime. In contrast, disc-dominated galaxies exhibit a significant wavelength dependence over the entire wavelength range considered. Their concentration is typically below 2.5 in the UV range, because of the widespread contribution of young stars across the disc. As the wavelength shifts to the optical, the concentration increases sharply, driven by the prominence of the central bulge \citep[e.g.,][]{Nersesian2023}. The wavelength dependence of the concentration index tends to flatten in the NIR, with median values that are still 0.5 or more below the one for bulge-dominated galaxies. It is consistent with the SINGS results of \citet{MunozMateos2009}, who note a small decline when moving toward the NIR. Our results are also in line with those of \citet{Baes2020c}, \citet{Nersesian2023}, and \citet{Kapoor2021}. 
 
An interesting wavelength dependence of the Gini coefficient can be seen in the top-right panel of Fig.~\ref{fig:figure_4}. Contrary to the concentration, the Gini coefficient has a non-monotonous wavelength dependence in the UV-to-NIR range: it peaks in the UV, decreases toward optical wavelengths, and then gradually increases again from optical to NIR. The same qualitative behaviour is found for both the disc-dominated and bulge-dominated populations, with the latter systematically having higher Gini values at all wavelengths. The origin of the high Gini index in the UV is bright emission from clumpy star-forming regions, which stand out against the relatively low-level continuum from older stellar populations. In the optical, the contrast between the star-forming regions and the older populations decreases, whereas in the NIR the bright bulge is responsible for the increase in Gini coefficient. The SINGS galaxies from \citet{MunozMateos2009} show a similar qualitative behaviour, but we note an offset between the curves. This offset could be related to the aperture within which the Gini index is calculated. The values for the SINGS sample are standardly calculated within the $R_{25}$ elliptical aperture. If the Petrosian radius $r_{\text{p}}$ is used instead of the $R_{25}$ radius, the Gini index decreases: \cite{MunozMateos2009} found that, on average, $\langle G_{R_{25}} - G_{r_{\text{p}}} \rangle = 0.11$, with a typical scatter of $\pm 0.10$. Accounting for this offset, the SINGS values fully agree with our TNG50 results (dotted orange line).

As shown in the bottom-left panel of Fig.~\ref{fig:figure_4}, the $M_{20}$ parameter exhibits a strong dependence on wavelength. In the UV bands, TNG50 galaxies show relatively high (less negative) $M_{20}$ values, indicating a more extended light distribution dominated by bright star-forming regions. From the UV to the optical, $M_{20}$ decreases steadily, becoming more negative. This trend suggests a more centrally concentrated and symmetric flux distribution at longer wavelengths. In the NIR, $M_{20}$ shows minimal variation, remaining nearly constant. This wavelength dependence is broadly consistent with the SINGS measurements by \citet{MunozMateos2009} and with the results by \citet{Baes2020c}. Disc-dominated galaxies tend to have higher (less negative) $M_{20}$ values at all wavelengths, due to their extended star-forming discs. In contrast, bulge-dominated galaxies exhibit more negative $M_{20}$ values, reflecting their more concentrated light profiles. For both types, the transition to more negative $M_{20}$ values with increasing wavelength reflects the reduced contribution of clumpy star-forming regions and the growing dominance of older stellar populations. 

Finally, the bottom-right panel of Figure~\ref{fig:figure_4} illustrates how the bulge statistic $F$ varies with wavelength. At shorter wavelengths, where emission is dominated by young, star-forming regions, the bulge appears less prominent, resulting in lower $F$ values. Conversely, at longer wavelengths, where older, redder stars dominate and dust attenuation is reduced, the bulge becomes more visible, leading to higher $F$ values. Disc-dominated and bulge-dominated galaxies show a similar wavelength dependence, but with a clear offset of about 0.4 across the entire UV--NIR wavelength range. Interestingly, bulge-dominated galaxies appear on average disc-dominated in the UV ($\langle F_{\text{UV}} \rangle < 0$), whereas disc-dominated galaxies appear bulge-dominated in the NIR ($\langle F_{\text{NIR}} \rangle > 0$). The comparison with the SINGS measurements is satisfactory, at least when we use the estimate based on the Petrosian aperture.

In summary, we find a clear wavelength dependence of all morphological indicators and a systematic difference between the disc-dominated and bulge-dominated galaxy populations. In general, galaxies become more concentrated when we move from UV to NIR wavelengths. This reflects the diminishing influence of star-forming regions and the growing dominance of older and redder stellar populations when wavelength increases.

%%%%%%%%%%%%%%%%%%%%%%%%%%%%%%%%%%%%%%%%%%%%%%%%%%%%%
 
\subsection{Wavelength dependence of the distribution in the Gini--$M_{20}$ plane}
\label{subsec:gini_m20_vs_lambda}

\begin{figure*}[t]
\centering
\includegraphics[width=\textwidth]{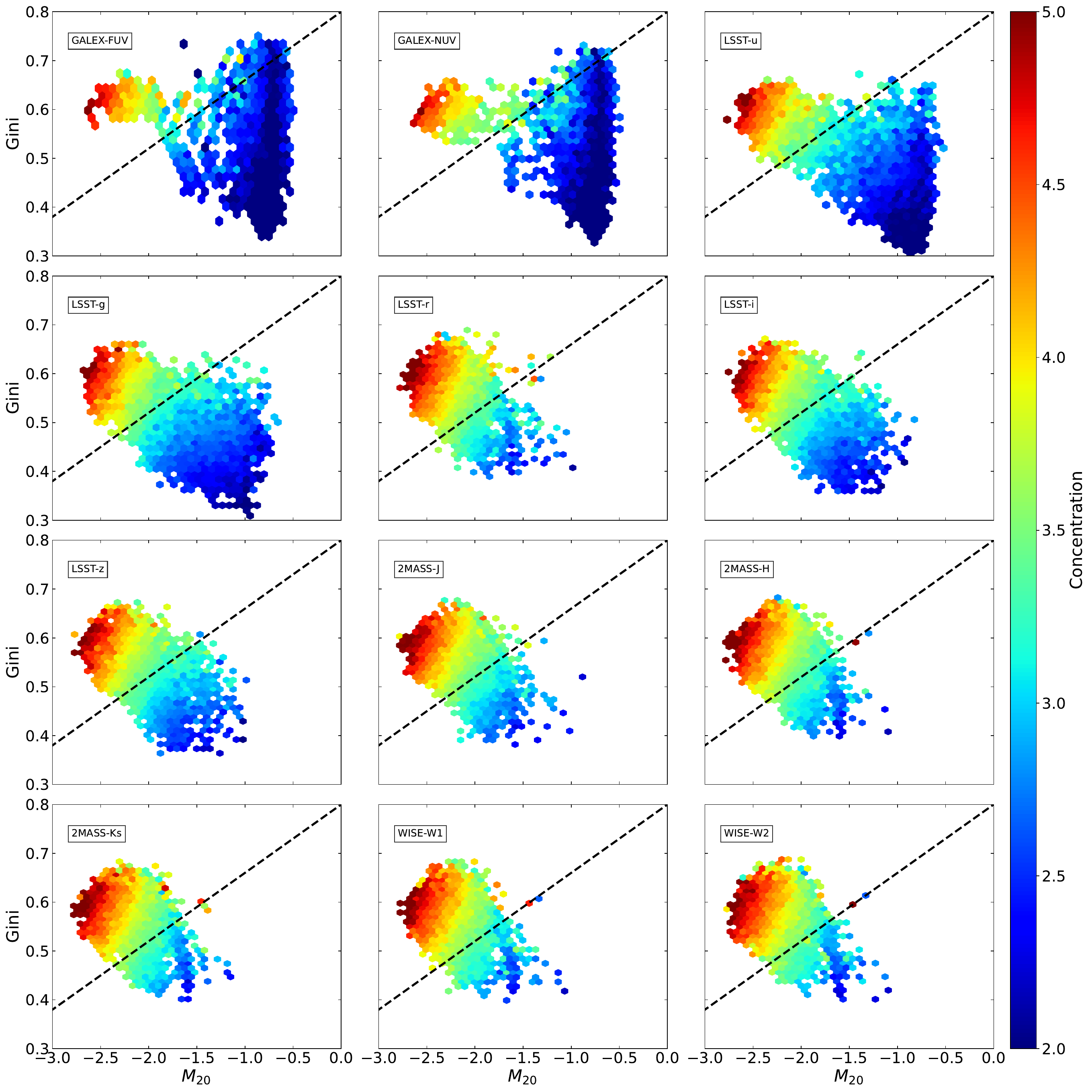}
\caption{
Wavelength dependence of the distribution in the Gini--$M_{20}$ plane. This multi-panel figure shows how galaxy morphology, as measured by the Gini coefficient (\textit{y}-axis) and $M_{20}$ (\textit{x}-axis), varies across 12 different bands from UV to NIR. Each panel corresponds to a different filter and is colour-coded according to the concentration index, a measure of how centrally the light is concentrated. A black dashed line in each panel shows the Gini--$M_{20}$ demarcation from \citep{Lotz2008}, which divides bulge-dominated galaxies above the line from disc-dominated (disc-like or irregular) galaxies below it.}
\label{fig:figure_5}
\end{figure*}

The second panel of Figure~{\ref{fig:figure_1}} showed the $i$-band Gini--$M_{20}$ plane for our TNG50 galaxies. Figure~\ref{fig:figure_5} extends this plot and presents the wavelength dependence of the distribution in the Gini--$M_{20}$ plane in each of the 12 individual bands considered, showcasing how galaxy morphology varies from UV to NIR wavelengths. Each panel contains measurements for each galaxy in our sample, in each of the four orientations considered (O1--O4). The morphological parameters are computed from the dust-attenuated synthetic images of the TNG50-SKIRT Atlas described in Sect.~{\ref{sec:methods}}, including the effects of instrumental resolution and background noise.

In the UV, most TNG50 galaxies are located in the disc-dominated region of the Gini--$M_{20}$ diagram. UV wavelengths are dominated by the emission from clumpy star-forming regions that dominate disc structures. Only a relatively small population of bulge-dominated galaxies with little or no ongoing star formation is located in the bulge-dominated region. As the wavelength increases, galaxies gradually shift toward the bulge-dominated region. These trends highlight that the old stellar populations, which are well traced at longer wavelengths, are more centrally concentrated, suggesting that the stellar mass profile tends to be bulge-dominated. This is consistent with previous findings that galaxy morphologies appear more bulge-like at redder wavelengths \citep{Wuyts2012, Ren2024}. Importantly, this demonstrates that nonparametric morphological classifications are sensitive to the wavelength of observation. As a result, morphology classifications can be biased if wavelength effects are not taken into account. Correcting for this bias through morphological k-corrections is essential, as noted in \citet{TaylorMager2007} and \citet{Yao2023}.

\subsection{Optimal bands to characterise galaxy morphology}

The wavelength dependence of the nonparametric morphology indicators also provides guidance on the optimal choice of photometric bands for morphological studies. UV bands are strongly dominated by emission from young stellar populations and are therefore highly sensitive to recent star formation and dust attenuation, resulting in clumpy and irregular morphologies that are not representative of the global stellar structure of galaxies. At the other extreme, NIR bands trace the bulk of the stellar mass and are much less affected by dust attenuation, yielding smoother and more stable morphological measurements that more closely reflect the intrinsic stellar structure of galaxies.

Optical bands (such as {\em{g}}, {\em{r}}, or {\em{i}}) provide an intermediate regime, combining sensitivity to both young and old stellar populations while still being affected by dust attenuation at a moderate level. As such, they offer a useful compromise for morphological classification, capturing the main structural components of galaxies while retaining sensitivity to features such as spiral arms and dust lanes. 

From an observational perspective, optical bands also benefit from the wide availability of high-quality imaging data from current and upcoming surveys, such as DESI \citep{DESICollaboration2025} and the Rubin Observatory \citep{Ivezic2019}, while NIR coverage will be significantly expanded by missions such as \textit{Euclid} \citep{EuclidCollaborationMellier2025} and the Nancy Grace Roman Telescope \citep{Spergel2015}. In contrast, UV imaging remains comparatively limited in sky coverage and depth.

The optimal choice of wavelength therefore depends on the scientific objective. For studies aiming to recover the intrinsic stellar structure of galaxies, NIR bands are preferred. For comparisons with large observational surveys and for morphological classification in a traditional sense, optical bands provide a robust and widely applicable choice. UV bands, on the other hand, are more suitable for studies of star formation morphology rather than global structural classification.

%%%%%%%%%%%%%%%%%%%%%%%%%%%%%%%%%%%%%%%%%%%%%%%%%%%%%

\section{Effect of dust attenuation on galaxy morphology}
\label{sec:dust_effect_on_npd}

\begin{figure*}[t]
\centering
\includegraphics[width=0.8\textwidth]{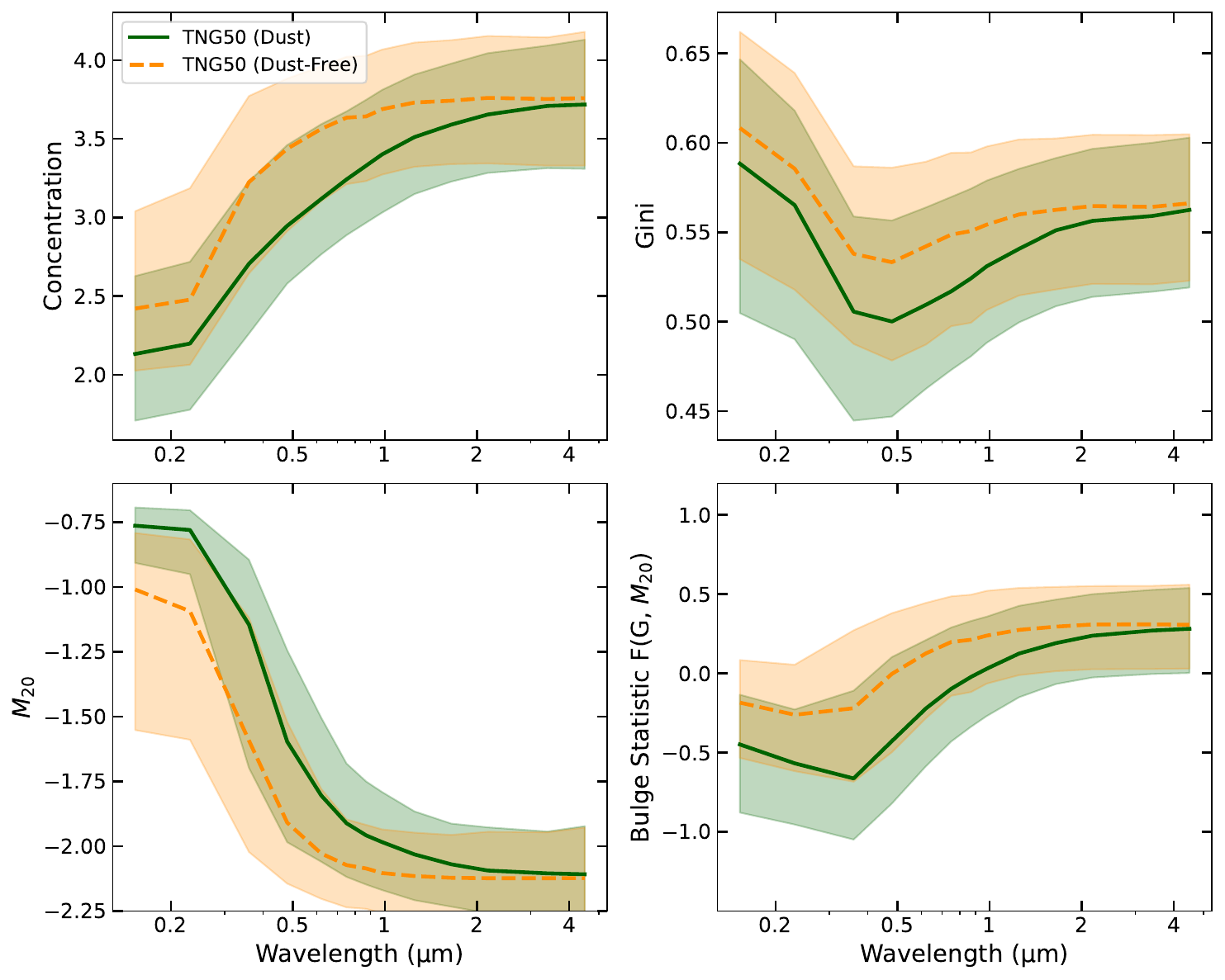}
\hspace*{2em}
\caption{Dependence of nonparametric morphology indicators on dust attenuation. The solid and dotted lines show the nonparametric indicators corresponding to dust-attenuated and dust-free images, respectively. The shaded regions represent the $25\%$ to $75\%$ percentile ranges.}
\label{fig:figure_6}
\end{figure*}

Previous radiative transfer studies have explored the impact of dust on parametric morphology using idealized galaxy models \citep{Graham2008, Gadotti2010, Pastrav2013a, Pastrav2013b}. \citet{RodriguezGomez2019} and \citet{GuzmanOrtega2023} studied the effect of dust attenuation on morphology, focusing on a single broadband filter ({\it{i}}-band and {\it{r}}-band, respectively). In both cases, they did not find a significant impact of dust attenuation on morphology. The influence of dust attenuation on nonparametric morphology indicators remains largely unexplored. The TNG50-SKIRT Atlas is ideal for this purpose, as it contains both dust-attenuated and dust-free images for all galaxies in all bands. We hence can directly investigate the effect of dust attenuation on morphological measurements in any band for individual galaxies or for the population as a whole. A similar approach has been applied to study the effect of dust attenuation on the wavelength dependence of the effective radius of the galaxy in the TNG50-SKIRT Atlas \citep{Baes2024b}.

To quantify the level of dust attenuation in our sample, we calculate the mean {\it{V}}-band attenuation for each galaxy by computing the difference between the dust-attenuated and dust-free magnitudes, averaged over all viewing orientations. We find a median attenuation of approximately 0.20 mag, with a 16th–84th percentile range of 0.03--0.45 mag across the sample. Disc-dominated galaxies exhibit higher attenuation ($\sim$0.25 mag) compared to bulge-dominated systems ($\sim$0.15 mag). Overall, dust attenuation is modest, but it shows a clear dependence on galaxy structure, with higher values in disc-dominated systems, consistent with the increase in dust attenuation with disc inclination.

\subsection{Median population trends}

Fig.~\ref{fig:figure_6} compares the wavelength dependence of the median values of the four nonparametric morphology indicators for dust-attenuated and dust-free galaxy images. We observe that dust attenuation generally affects various nonparametric morphology indicators. These effects are most significant at the shortest wavelengths, where the efficiency of absorption and scattering by dust is highest. In the NIR, where dust attenuation is least efficient, the measurements of dust-attenuated and dust-free morphological parameters converge. 

As expected, the top-left panel of Fig.~{\ref{fig:figure_6}} shows that dust-attenuated galaxies exhibit, on average, lower concentration values compared to dust-free galaxies. The removal of light by absorption and scattering from the central region, combined with scattering into the line of sight at larger radii, makes galaxies appear less concentrated in their central regions when observed in the UV--optical regime \citep{Calzetti2001, Lotz2008}. This trend is consistent with the increase in the effective radius of galaxies as a result of dust attenuation \citep{Kelvin2012, Vulcani2014, Nedkova2024, Baes2024b}.

The remaining three panels of Fig.~{\ref{fig:figure_6}} show how dust attenuation affects the average position of galaxies in the Gini--$M_{20}$ plane. Dust attenuation tends to decrease the Gini index and increase $M_{20}$ for the same reason as discussed above: dust in galaxies is preferentially concentrated toward the central regions, where it efficiently absorbs a significant fraction of UV–optical radiation. The net effect is that the central regions of galaxies are more affected by dust than the outskirts, which, conversely, are barely sensitive to the presence of dust. The combination of these two effects automatically leads to a decrease in the bulge statistic as a result of dust attenuation. In other words, dust attenuation tends to make galaxies look more disc-like than they would be if there were no dust in the systems. This is completely in agreement with the results of \citet{Gadotti2010}, who found that dust attenuation leads to a strong underestimate of the bulge-to-disc ratio in galaxies.

\subsection{Dust attenuation effects in the Gini--$M_{20}$ plane}

\begin{figure*}[t]
\centering
\includegraphics[width=\textwidth]{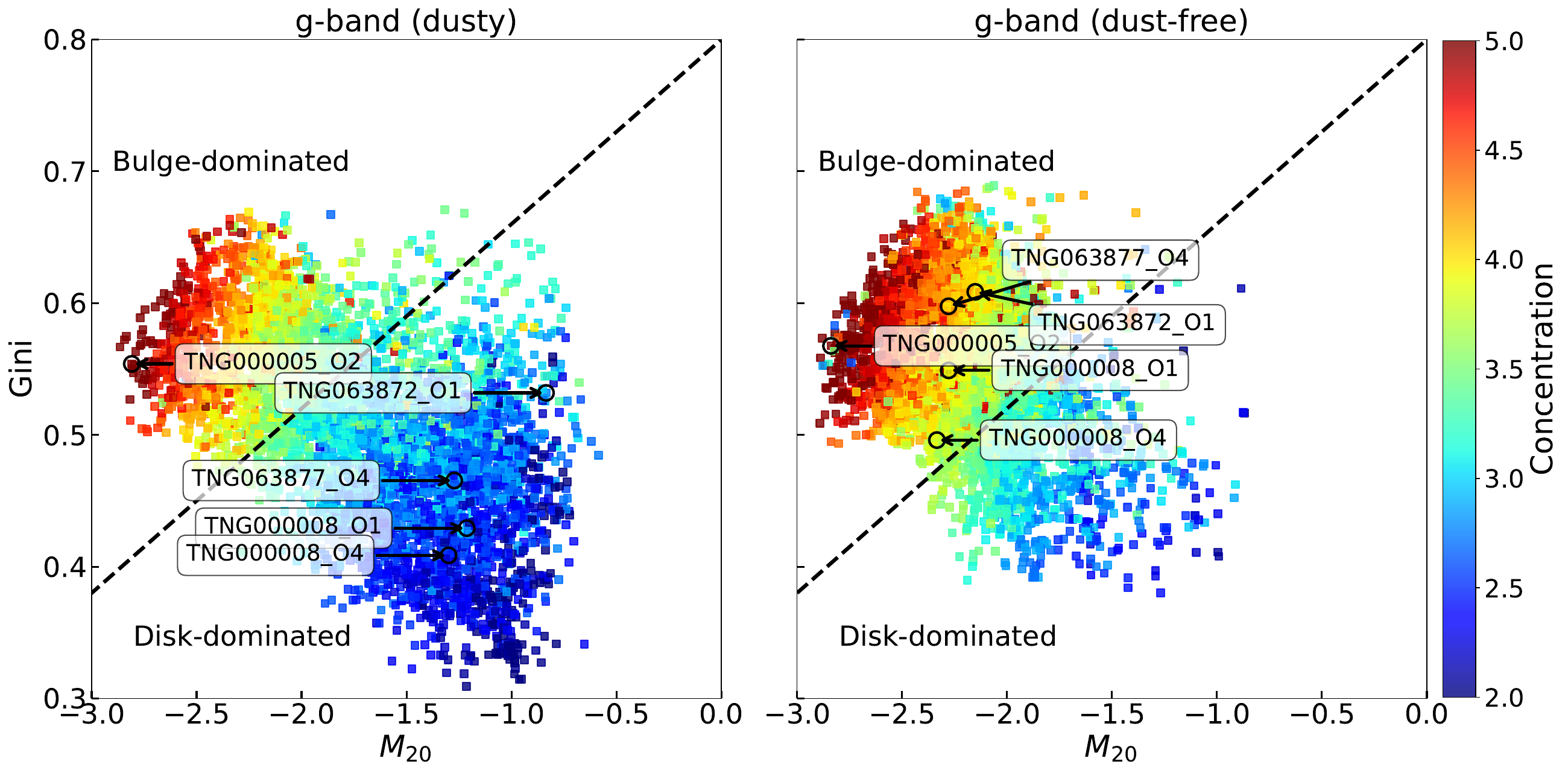}
\caption{Examples on which the effect of dust attenuation is significant in some galaxies in the Gini-$M_{20}$ diagram. The left panel shows Gini-$M_{20}$ diagram for dust-attenuated $g$-band, whereas right hand panel illustrates dust-free $g$-band galaxies. Dust typically lowers concentration, Gini and increases $M_{20}$ values, shifting galaxies toward disc-dominated classifications. The black dotted division line \citep{Lotz2008} separates bulge-dominated galaxies from disc-dominated galaxies.}
\label{fig:figure_7}
\end{figure*}

Figure~{\ref{fig:figure_7}} illustrates the distribution of TNG50 galaxies in the {\it g}-band Gini--$M_{20}$ plane, comparing dust-attenuated and dust-free images. At the population level, dust attenuation shifts the galaxy distribution toward lower Gini and concentration values, while $M_{20}$ becomes slightly higher (i.e. less negative). This behaviour is consistent with the physical effect of dust attenuating and redistributing central light. In the absence of dust, galaxies tend to occupy regions with higher Gini and concentration values and lower $M_{20}$, reflecting a more centrally concentrated light distribution.

However, these shifts are not uniform on a galaxy-by-galaxy basis. While some galaxies exhibit negligible changes due to their low dust content, others show more substantial variations in their nonparametric morphology indicators. The impact of dust also depends on galaxy structure: disc-dominated systems typically show lower Gini and concentration values and slightly higher $M_{20}$ when dust is included, whereas bulge-dominated galaxies, which are more centrally concentrated, are less affected by dust attenuation.

\begin{figure*}[t]
\centering
\includegraphics[width=\textwidth]{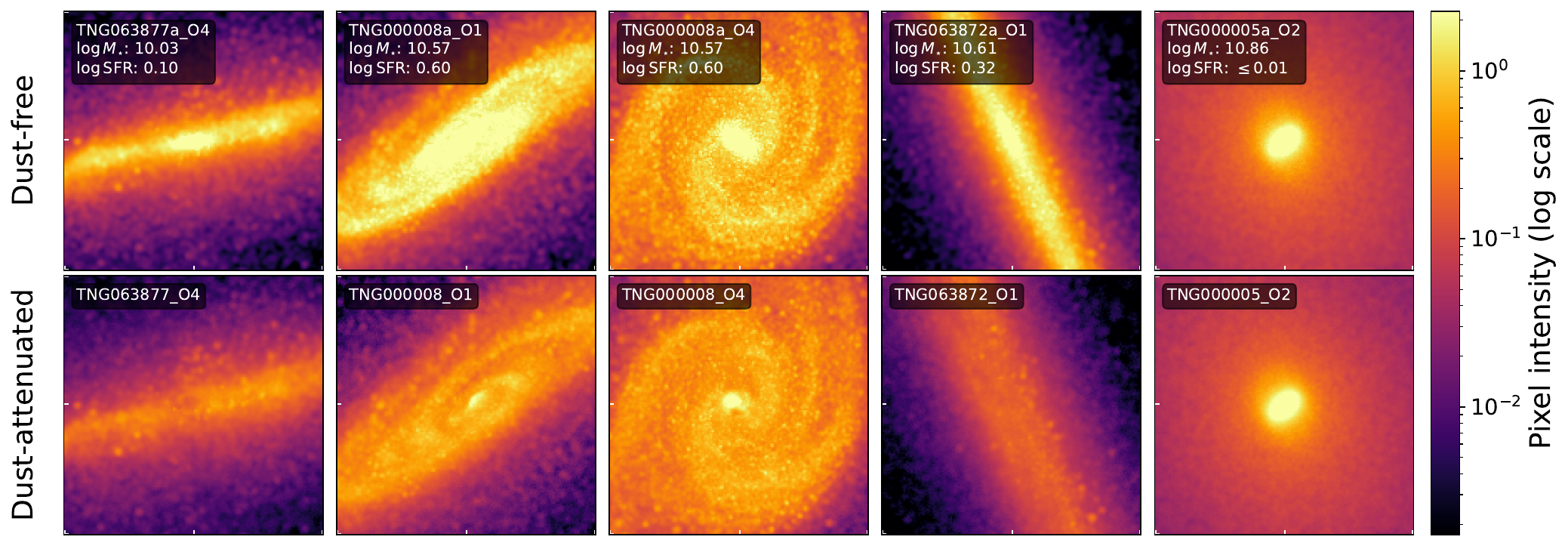}
\caption{
Dust-free (top row) and dust-attenuated (bottom-row) {\it{g}}-band images for five TNG50-SKIRT Atlas galaxies, selected to have the largest morphological shifts of Gini and $M_{20}$ due to dust attenuation. Each images is a cutout with a field-of-view of 30~kpc.}
\label{fig:figure_8}
\end{figure*}

Five representative examples in our sample are indicated in 
Figure~\ref{fig:figure_7}:
\begin{itemize}
    \item TNG\,000005\_O2 is a massive elliptical galaxy with no dust and no ongoing star formation. Its Gini and $M_{20}$ indices are therefore unaffected by dust attenuation.     
    \item TNG\,000008\_O1 and TNG\,063877\_O4 are the two galaxies that exhibit the largest shifts in the Gini index due to dust attenuation in the sample. They are both moderately inclined spiral galaxies.
    \item TNG\,063872\_O1 and TNG\,000008\_O4 are two edge-on spiral galaxies that suffer significant dust attenuation. They show the largest shifts in $M_{20}$.
\end{itemize}
Figures~{\ref{fig:figure_7}} and {\ref{fig:figure_8}} illustrate the effect of dust attenuation for these five representative galaxies. Figure~{\ref{fig:figure_7}} shows their positions in the $g$-band Gini--$M_{20}$ plane, while Fig.~{\ref{fig:figure_8}} presents the corresponding dust-attenuated and dust-free image cutouts. Together, these figures highlight how dust alters morphological indicator values at the level of individual galaxies.

%%%%%%%%%%%%%%%%%%%%%%%%%%%%%%%%%%%%%%%%%%%%%%%%%%%%%

\section{Conclusions}
\label{sec:summary_and_conclusions}

\subsection{Summary and conclusions}

In this paper, we have studied the wavelength dependence of galaxy morphology for a sample of TNG50 galaxies at $z = 0$. The morphology was characterised using nonparametric morphology indicies (concentration, Gini coefficient, $M_{20}$, and bulge statistic), calculated from synthetic images in 12 broadband filters covering the UV to NIR wavelength range. The availability of multi-wavelength dust-attenuated and dust-free galaxies allows for a systematic study of the wavelength dependence of galaxy morphology and the effects of dust attenuation on morphological parameters. The large volume of TNG50 paired with its high resolution makes this a well-suited simulation for studying the detailed morphologies of the galaxy population. This results in a TNG50 simulation that is particularly special for studying the detailed morphologies of galaxies. All nonparametric morphology indicators of simulated galaxy images were fitted in a very similar manner using the {\tt StatMorph} code \citep{RodriguezGomez2019}, which calculates nonparametric morphological diagnostics such as Gini--$M_{20}$ \citep{Lotz2004} and CAS \citep{Conselice2003} statistics. We have compared the TNG50 morphologies with those of galaxies from the observational Pan-STARRS survey and from other hydrodynamical simulations, TNG100 and the original Illustris. We have also investigated the wavelength dependence of nonparametric morphology indicators for the entire galaxy population, classified into bulge- and disc-dominated systems, and compared these results with the observational SINGS sample. 

\begin{itemize}

\item We find that the locus of the Gini--$M_{20}$ diagram in TNG50 galaxies is consistent with the observation of Pan-STARRS and TNG100 simulated galaxies. The distributions agree well, despite the simulated galaxies being post-processed as isolated systems and, therefore, lacking a merging population. Furthermore, the distribution of galaxies in the concentration-Gini coefficient and concentration-$M_{20}$ diagrams for TNG50 are in good agreement with the TNG100 and Pan-STARRS samples.
\item Nonparametric morphological indicators change across the wavelength range, especially variation at the shorter wavelengths is more significant in all nonparametric morphology indicators. Our results show that TNG50 galaxies exhibit lower concentration and higher $M_{20}$ values in the UV than in the optical as a result of the UV tracing recent star formation, which has a clumpier and more radially extended spatial distribution. However, these galaxies show a flux distribution concentrated in fewer pixels and bulge-dominated structure at NIR wavelengths, which traces older stellar populations. Nonparametric morphology indicators vary significantly in UV-optical, but show only mild wavelength dependence in the NIR.
\item 
We find that disc-dominated galaxies appear to be less concentrated, have a lower Gini coefficient, have a higher $M_{20}$, and have a lower bulge statistic at shorter than at longer wavelengths.
\item 
Changes in concentration and $M_{20}$ with wavelength of TNG50 galaxies show relative consistency with observational samples. The Gini coefficient and bulge statistic of TNG50 galaxies show better agreement with the Gini coefficient and bulge statistic measured in the Petrosian aperture than in the $R_{25}$ aperture for SINGS galaxies.
\item 
We find that the effect of dust attenuation on nonparametric morphology indicators across all galaxy samples is modest. However, its effect is not minimal at the individual galaxy image level. Dust attenuation affects disc-dominated (spiral) galaxies significantly more than bulge-dominated (elliptical) galaxies, where it decreases concentration and Gini, and makes $M_{20}$ less negative, particularly in the UV and optical wavelength bands, but has little impact in the longer wavelength range.
\end{itemize}

\subsection{Caveats and outlook}

This work highlights the power of post-processed images from high-resolution simulations to disentangle the effects of stellar populations and dust attenuation on galaxy morphology. Nevertheless, the present analysis is subject to several limitations that should be kept in mind.
\begin{itemize}
\item 
AGN emission can in principle affect nonparametric morphology indicators by enhancing central light concentrations \citep{GetachewWoreta2022, GetachewWoreta2025}. However, the synthetic images used in this work are generated with the {\tt SKIRT} radiative transfer code and include only stellar and dust emission, without any explicit contribution from AGN. The overall contribution of optically bright AGN to the low-redshift galaxy population is known to be modest \citep{Goulding2009, Driver2018, Driver2026}. We identify AGN in our TSA sample based on the Eddington ratio of the central supermassive black hole. Using a conservative threshold of $\lambda_{\text{Edd}} > 0.01$, we find that only 3.9\% of galaxies in our sample host actively accreting black holes. No single galaxy has $\lambda_{\text{Edd}} > 0.1$, which would represent an optically bright AGN. This confirms that AGN are rare in the present sample and are unlikely to significantly affect the statistical trends discussed in this work. Any potential AGN influence is expected to primarily arise through its impact on galaxy evolution, rather than through a direct contribution of AGN light to the surface-brightness distribution. 
\item 
The synthetic images used in this work are affected by Monte Carlo noise inherent to the radiative transfer calculations. While this does not significantly impact the nonparametric indicators analysed in this study ($C$, $G$, $M_{20}$, and $F$), it can affect more noise-sensitive quantities such as asymmetry and smoothness, which were therefore excluded from the analysis. Future work will benefit from improved data products with reduced Monte Carlo noise, for example through the use of image-generation techniques optimised for suppressing noise in low-surface-brightness regions \citep{Baes2025}.
\item 
This study is limited to the UV--NIR wavelength range. However, longer wavelengths in the mid-infrared to mm regime also contain valuable morphological information, particularly related to dust emission and the interstellar medium \citep{MunozMateos2009, Baes2020c}. Such data can provide additional constraints on galaxy structure and evolution \citep{Kapoor2021, Camps2022}. Future releases of the TNG50-SKIRT Atlas will extend the wavelength coverage to the mm regime, enabling a more comprehensive multi-wavelength morphological analysis.
\end{itemize}

%%%%%%%%%%%%%%%%%%%%%%%%%%%%%%%%%%%%%%%%%%%%%%%%%%%%%

\begin{acknowledgements}
SBT and ATE appreciate financial support from the NASCERE project, a bilateral cooperation program between Jimma University (Ethiopia) and Ghent University (Belgium). MB and MM acknowledge financial support from the Flemish Fund for Scientific Research (FWO-Vlaanderen) through the research project G030319N. MB and AN acknowledge support from the Belgian Science Policy Office (BELSPO) through the PRODEX project ``Belgian Euclid Science Exploitation (BESE)'' (No. 4000143202).
\end{acknowledgements}

%%%%%%%%%%%%%%%%%%%%%%%%%%%%%%%%%%%%%%%%%%%%%%%%%%%%%

\bibliography{mybib}

%%%%%%%%%%%%%%%%%%%%%%%%%%%%%%%%%%%%%%%%%%%%%%%%%%%%%

\end{document}